\documentclass[11pt]{article}
\usepackage[margin=1in]{geometry}
\usepackage[T1]{fontenc}
\usepackage[utf8]{inputenc}
\usepackage{amsmath,amssymb,amsfonts}
\usepackage{graphicx}
\usepackage{cite}
\usepackage{url}
\usepackage{array}
\usepackage{xcolor}
\usepackage[colorlinks=true,linkcolor=blue,citecolor=blue,urlcolor=blue,hypertexnames=false]{hyperref}
\DeclareUnicodeCharacter{03B3}{$\gamma$}
\DeclareUnicodeCharacter{03BC}{$\mu$}
\newcommand{\mus}{\,\mu\mathrm{s}}
\newcommand{\figref}[1]{Fig.~\ref{#1}}
\newcommand{\figsubref}[2]{\hyperref[#1]{Fig.~\ref*{#1}(#2)}}
\begin{document}
\setlength{\emergencystretch}{3em}
\sloppy

\title{Single-Pulse Dosimetry for Ultra-High-Dose-Rate X-Ray Pulses Based on Radiation-Induced Acoustic Signals}
\date{}

\author{\small Qihang Han, Ke Chen, Chenghao Hua, Haoran Chen, Lili Ding, Changhui Li, Xiaoping Ouyang, Chenhui Wang,\\ Wei Chen, and Wenjun Ma}

\maketitle
\begin{center}
\footnotesize Qihang Han, Ke Chen, Chenghao Hua, and Haoran Chen: State Key Laboratory of Nuclear Physics and Technology, School of Physics, Peking University, Beijing 100871, China.
Chenhui Wang, Wei Chen, and Lili Ding: National Key Laboratory of Intense Pulsed Radiation Simulation and Effect, Northwest Institute of Nuclear Technology, Xi'an 710024, China.
Xiaoping Ouyang: National Key Laboratory of Intense Pulsed Radiation Simulation and Effect, Northwest Institute of Nuclear Technology, P.O. Box 69-10, Xi'an 710024, China.
Wenjun Ma: State Key Laboratory of Nuclear Physics and Technology, School of Physics, Peking University, Beijing 100871, China, and Beijing Laser Acceleration Innovation Center, Huairou, Beijing 101400, China.
Changhui Li: Department of Biomedical Engineering, College of Future Technology, Peking University, Beijing 100871, China.
Corresponding authors: Chenhui Wang (0310wch@pku.edu.cn), Wei Chen (chenwei6802@163.com), and Wenjun Ma (wenjun.ma@pku.edu.cn).
\end{center}
\textcopyright\ 2026 IEEE. Personal use of this material is permitted. Permission from IEEE must be obtained for all other uses, including reprinting or republishing this material for advertising or promotional purposes, collecting new collected works for resale or redistribution to servers or lists, or reuse of any copyrighted component of this work in other works.

\begin{abstract}
\boldmath
Real-time dosimetry of ultra-high-dose-rate (UHDR) pulsed X-rays remains a major challenge. This proof-of-principle study investigates a dosimetry method for UHDR pulsed X-rays based on radiation-induced acoustic (RA) signals from a thin metal foil. The foil combines efficient X-ray absorption and acoustic resonance, converting deposited dose into acoustic signals with discrete resonant frequencies. Single-pulse measurements were performed at six dose levels spanning approximately 1--40 mGy(Si), with one pulse recorded at each level and instantaneous dose rates reaching the order of $10^{6}\,\mathrm{Gy/s}$ at the highest dose settings. The peak-to-peak amplified transducer-output voltage exhibited a linear relationship with the reference dose measured by thermoluminescent dosimeters (TLDs). These results demonstrate the feasibility of single-pulse acoustic dosimetry under the tested conditions. Furthermore, we establish analytical solutions for radiation-induced acoustic signals in a one-dimensional layered medium in both the time and frequency domains, providing a theoretical basis for signal interpretation and detector optimization. 
\end{abstract}


\section{Introduction}

Recent advances in electron induction accelerators\cite{ref1}, radio-frequency linear accelerators\cite{ref2}, and high-current pulsed-power systems\cite{ref3,ref4} have enabled intense pulsed radiation sources capable of producing nanosecond X/$\gamma$-ray pulses. Such sources typically deliver $10$--$100\,\mathrm{ns}$ pulses at ultra-high instantaneous dose rates (UHDR) on the order of $10^{6}\,\mathrm{Gy/s}$ and are widely used for diagnostics in high-energy-density physics (HEDP)\cite{ref5}, flash radiography, and radiation-effects testing\cite{ref6}. In the aerospace and electronics industries, UHDR radiation is also used to evaluate the radiation tolerance of critical components under extreme conditions. Shot-to-shot real-time monitoring of the dose per pulse is critical not only for maintaining stable radiation output and rapidly detecting emerging issues, but also for improving experimental efficiency.

However, real-time single-shot X-ray dosimetry remains a major challenge in the UHDR regime. The performance of most commercially available real-time dosimeters deteriorates severely at UHDR, with dose-rate dependence, signal saturation, and a strongly nonlinear response\cite{ref7,ref8,ref9}. Currently, dose monitoring at UHDR mainly relies on passive, offline dosimeters such as radiochromic films (RCF)\cite{ref10}, thermoluminescent dosimeters (TLDs)\cite{ref11}, and alanine dosimeters\cite{ref12}. These passive dosimeters provide reference measurements but require post-irradiation readout. Online dosimetry mainly uses charge-collection detectors, such as ionization chambers and semiconductor devices, and optical approaches based on Cherenkov emission or scintillation. Ionization chambers are often used as reference detectors for their high accuracy. However, at UHDR, the extremely high ionization density increases charge recombination, which reduces collection efficiency and can saturate the measured current, ultimately complicating corrections and increasing uncertainty. This effect becomes particularly pronounced when the instantaneous dose rate exceeds approximately $10^{6}\,\mathrm{Gy/s}$. Although specialized chambers such as the Advanced Markus ionization chamber have demonstrated improved tolerance and can provide accurate measurements up to $5.56\,\mathrm{mGy}$ per pulse at dose rates approaching $10^{6}\,\mathrm{Gy/s}$, they remain susceptible to saturation and response degradation at higher intensities\cite{ref13}. Similarly, semiconductor detectors offer high sensitivity but often show dose-rate-dependent nonlinearity in the UHDR regime\cite{ref14}. Optical approaches based on Cherenkov emission or scintillation enable real-time, spatially resolved monitoring\cite{ref15,ref16}. However, the energy threshold and spectral dependence of Cherenkov readout, together with possible Cherenkov contamination of scintillation signals, complicate quantitative calibration.

A promising approach for pulsed X-ray dosimetry exploits the radiation-induced acoustic (RA) effect, in which the absorbed dose is inferred from acoustic pressure generated by rapid energy deposition\cite{ref17,ref18,ref19,ref20}. This effect was first demonstrated in 1991 by Bowen et al., who recorded acoustic signals generated in water by therapeutic X-ray irradiation\cite{ref21}. RA occurs when X-ray pulses interact with a material, exciting electrons that transfer kinetic energy to the material via electron-phonon coupling, causing localized thermoelastic expansion and generating pressure waves\cite{ref22,ref23,ref24,ref25}. In water, when energy from an infinitesimally short pulse is deposited into an infinitesimally small volume, the initial pressure generated by a $1\,\mathrm{Gy}$ dose at $37\,^{\circ}\mathrm{C}$ is estimated to be approximately $200\,\mathrm{Pa}$, which can be readily recorded using a piezoelectric transducer with a low-noise amplifier. Because its readout does not rely on charge collection, RA can mitigate recombination-induced nonlinearity and saturation in the UHDR regime. Taken together, these considerations make RA particularly well suited for quantitative dosimetry of UHDR pulses.

In this work, we present a resonant, foil-based radiation-induced acoustic detector for quantitative, shot-resolved dosimetry of UHDR X-ray pulses. The large acoustic-impedance mismatch at the lead-air and lead-water interfaces produces strong reflections that form an acoustic cavity. The resulting well-defined narrowband resonance and characteristic ring-down signature facilitate single-shot readout. Using UHDR X-ray pulses from the ``Chen Guang'' accelerator, we record single-pulse acoustic signals and establish the relationship between their peak-to-peak voltage and the TLD reference dose. To interpret the measurements quantitatively, we further develop a thermoacoustic model that relates the measured response to the depth-dependent energy-deposition profile within the foil. The main innovations of this study are summarized as follows.

\begin{enumerate}
\item Unlike existing online UHDR dosimetry techniques that rely on charge collection or optical readout\cite{ref11,ref14}, the proposed resonant acoustic approach employs a passive ultrasonic transducer, making it suitable for operation in complex radiation environments. The acoustic response is temporally separated from the prompt electromagnetic interference. In addition, the signal exhibits a resonant ring-down with a characteristic periodic structure, facilitating acoustic-response identification and extraction of amplitude features for dose analysis.
\item Previous X-ray radiation-induced acoustic studies were mainly performed with clinical linear accelerators under conventional radiotherapy conditions, typically at instantaneous dose rates below $10^{3}\,\mathrm{Gy/s}$ and with substantial signal averaging for dose imaging\cite{ref26,ref27,ref28}. In contrast, this work establishes a method for single-pulse dosimetry of UHDR pulsed X-rays. The experiments show a linear relationship between the measured acoustic response and the TLD reference dose, with instantaneous dose rates reaching the order of $10^{6}\,\mathrm{Gy/s}$ at the highest dose settings.
\item To the best of our knowledge, existing theoretical studies on resonant radiation-induced acoustics have mainly been limited to resonance-frequency estimation and finite-element simulations\cite{ref29}. In this work, we develop analytical solutions for radiation-induced acoustic waves in a one-dimensional layered medium in both the frequency and time domains. The model explicitly relates the depth-dose distribution and cavity resonance to the measured acoustic response, providing a physics-based framework for interpreting experimental observations and guiding detector design and optimization.
\end{enumerate}

\section{Resonant acoustic signals from the interfaces}

\begin{figure}[!t]
\centering
\includegraphics[width=\textwidth]{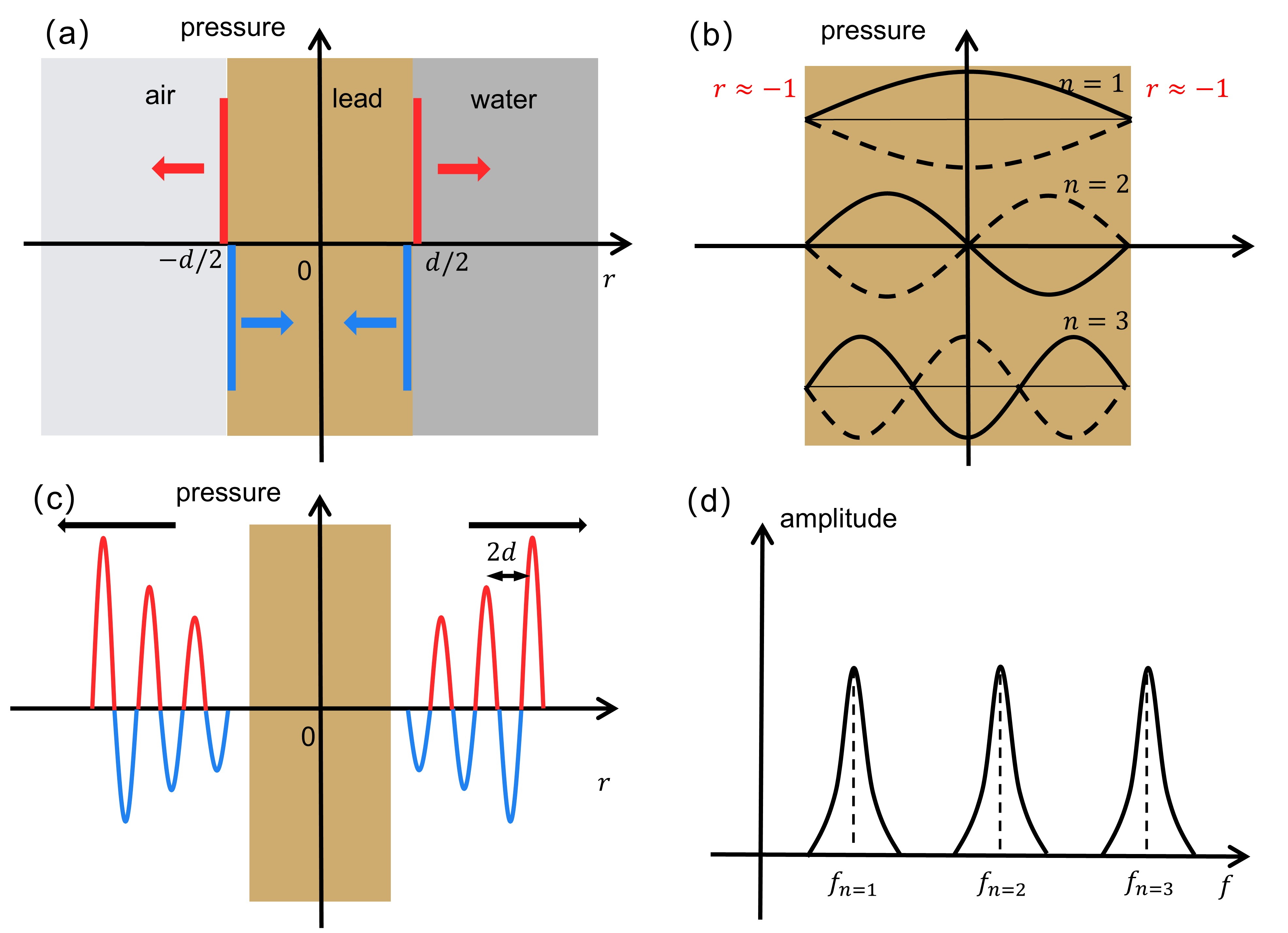}
\caption{Schematic illustration of the resonant acoustic detection mechanism and the resulting acoustic signal characteristics in both time and frequency domains.}
\label{fig:1}
\end{figure}

\figsubref{fig:1}{a} illustrates the core concept of the resonant radiation-induced acoustic detector. A lead foil separates air and water, with the X-ray pulse incident from the air side. Because the foil absorbs X-rays efficiently, the deposited energy density changes sharply across the lead-air and lead-water interfaces, producing steep energy density gradients. We assume laterally uniform irradiation and neglect depth variation in the deposited energy density across the foil. These gradients create initial pressure discontinuities that launch bipolar acoustic waves comprising outward-propagating compressional waves (red arrows) and inward-propagating rarefaction waves (blue arrows). The large acoustic impedance contrast between lead and the surrounding media (\mbox{$Z_{\mathrm{lead}}=25.65\,\mathrm{MRayl}$}, \mbox{$Z_{\mathrm{water}}=1.50\,\mathrm{MRayl}$}, and \mbox{$Z_{\mathrm{air}}=411\,\mathrm{Pa\,s/m}$}) produces strong reflections with pressure phase inversion. The pressure-reflection coefficient is close to \mbox{$-1$} at the Pb--air interface and approximately \mbox{$-0.89$} at the Pb--water interface, allowing partial transmission into water. The \mbox{$r\approx-1$} annotation in \mbox{\figsubref{fig:1}{b}} represents an idealized reflection condition. Under these conditions, repeated reflections produce frequency-dependent interference that preferentially enhances the thickness-resonance components in the transmitted spectrum. The transmitted signal therefore exhibits a resonant ring-down that can be detected by an ultrasound transducer.

\figsubref{fig:1}{c} schematically illustrates the emitted acoustic wave train in space, whereas \figsubref{fig:1}{d} depicts the corresponding resonant peaks in the frequency domain. In the time domain, the compressional wave reaches the lead-water interface and is partially transmitted into water, followed by a rarefaction wave from the opposite (lead-air) boundary. The superposition of these waves at the right interface generates the first bipolar (positive-negative) acoustic signal. The second bipolar signal is produced when the rarefaction wave propagates from the right boundary to the left, reflects at the left interface with phase inversion, and returns to the right as a compressional wave. The time delay between successive bipolar signals corresponds to the round-trip travel time of sound within the lead foil, consistent with standing wave patterns. In the frequency domain, the resonance condition results in discrete peaks, as depicted in \figsubref{fig:1}{d}. These resonances occur when the standing-wave condition \mbox{$f=nv_i/(2l)$} is satisfied, where \mbox{$n=1,2,3,\ldots$}, \mbox{$v_i$} is the sound velocity in lead, \mbox{$l$} is the layer thickness, and \mbox{$f$} is the acoustic frequency\cite{ref39}.

\section{Experimental setup and results}

\subsection{Experimental setup and measurement procedure}

\begin{figure}[!t]
\centering
\includegraphics[width=\linewidth]{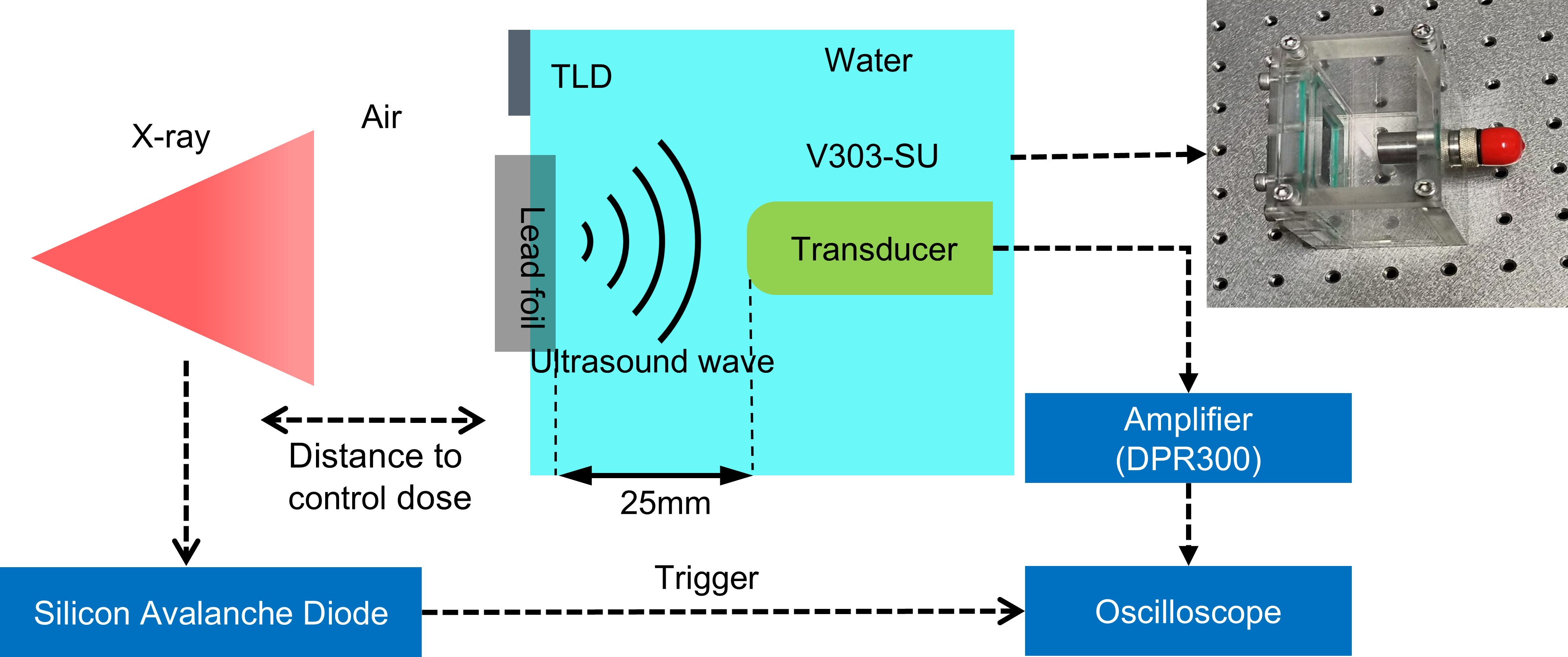}
\caption{Schematic of the radiation-induced acoustic dose-measurement apparatus and readout chain. The lead foil serves as the X-ray absorber and acoustic resonator, and the water coupling layer transmits acoustic waves to the piezoelectric receiving transducer. The transducer output is amplified and recorded by the oscilloscope. A silicon avalanche photodiode provides synchronous triggering, and a TLD provides an independent reference dose. The schematic is not to scale.}
\label{fig:2}
\end{figure}

\mbox{\figref{fig:2}} shows the experimental setup for radiation-induced acoustic measurements of UHDR X-ray pulses. A \mbox{$2\,\mathrm{mm}$}-thick lead foil serves as both the X-ray absorber and the acoustic resonator. Energy deposited by an X-ray pulse induces thermoelastic expansion of the foil, thereby generating acoustic waves that propagate through water to a piezoelectric ultrasound transducer. The transducer converts the acoustic waves into an electrical signal, which is amplified and recorded with an oscilloscope.

The lead foil is fixed at the front of the water chamber, forming the boundary between air and water. Its measured thickness was \mbox{$2.00\,\mathrm{mm}$}, and its longitudinal wave velocity, measured using the pulse-echo method, was \mbox{$2262\pm8\,\mathrm{m/s}$} (mean \mbox{$\pm$} sample standard deviation, \mbox{$n=3$}). An unfocused, circular immersion ultrasound transducer (V303-SU, Olympus-NDT, Waltham, MA, USA) is mounted on the opposite side of the chamber, facing the foil, with its acoustic axis aligned with the foil normal, at a water path length of \mbox{$25\,\mathrm{mm}$}. The lateral extent of the foil is substantially larger than the active aperture of the transducer; the acoustic field across the receiving aperture is therefore approximated as a plane wave at near-normal incidence. The transducer has a center frequency of \mbox{$1\,\mathrm{MHz}$}, a diameter of \mbox{$13\,\mathrm{mm}$}, a \mbox{$-6\,\mathrm{dB}$} bandwidth of 59\%, and an estimated receiving sensitivity of \mbox{$4.05\times10^{-6}\,\mathrm{V/Pa}$} near \mbox{$1\,\mathrm{MHz}$}. The \mbox{$2\,\mathrm{mm}$} foil thickness was selected to place the principal resonant components within the transducer's sensitive frequency range.

The transducer output was fed into a broadband, low-noise amplifier (DPR300, JSR Ultrasonics, Pittsford, NY, USA) with a gain of \mbox{$50\,\mathrm{dB}$} and a \mbox{$-3\,\mathrm{dB}$} bandwidth of \mbox{$1\,\mathrm{kHz}$}--\mbox{$3\,\mathrm{MHz}$}. Oscilloscope acquisition was synchronized with each X-ray pulse using an optical trigger provided by a silicon avalanche photodiode. The amplified voltage waveform from each pulse was recorded without waveform averaging at a sampling rate of \mbox{$100\,\mathrm{MS/s}$}, corresponding to a \mbox{$10\,\mathrm{ns}$} sampling interval, over a \mbox{$100\,\mu\mathrm{s}$} acquisition window. The RMS noise of the measurement system, including the transducer, amplifier, and oscilloscope, was \mbox{$8.44\,\mathrm{mV}$}.

The irradiation experiment was performed using UHDR pulsed X-rays generated by the ``Chen Guang'' accelerator through electron bremsstrahlung. The mean photon energy was \mbox{$200\,\mathrm{keV}$}, with a spectral endpoint of \mbox{$1\,\mathrm{MeV}$}. The pulse width was \mbox{$25\,\mathrm{ns}$}, with instantaneous dose rates reaching the order of \mbox{$10^{6}\,\mathrm{Gy/s}$} at the highest dose settings. X-rays were incident on the lead foil from the air side. Single-pulse doses of approximately \mbox{$1$}--\mbox{$40\,\mathrm{mGy}$} were obtained by translating the complete detector assembly along the beam axis to vary the source-to-detector distance. The accelerator operating settings and the relative positions of the foil and transducer remained unchanged throughout these measurements. The experiment was conducted under ambient laboratory conditions at approximately \mbox{$25\,^{\circ}\mathrm{C}$}.

GR-100M LiF(Mg,Ti) thermoluminescent dosimeters (TLDs), individually calibrated in a \mbox{$^{60}\mathrm{Co}$} reference field and read using an RGD-3A reader, provided independent references for the single-pulse dose measurements. For each measurement, a TLD was attached to the upper-central region of the lead foil and irradiated by the same X-ray pulse, with its position and orientation relative to the foil maintained at all dose settings. Each TLD was calibrated in terms of absorbed dose to silicon, and the resulting single-pulse reference dose, denoted by \mbox{$D_{\mathrm{ref}}$}, is reported in \mbox{$\mathrm{mGy(Si)}$}.

The uncertainty budget for the TLD reference dose is summarized in Table~\mbox{\ref{tab:tld_uncertainty}}, which lists the individual uncertainty contributions and the corresponding correction or control measures. Treating the components as uncorrelated, their combination by the root-sum-of-squares method gives a relative combined standard uncertainty of \mbox{$7.2\%$}. The corresponding relative expanded uncertainty is \mbox{$14.4\%$} for a coverage factor of \mbox{$k=2$}.

\begin{table}[!t]
\caption{Uncertainty budget for the TLD reference dose.}
\label{tab:tld_uncertainty}
\centering
\small
\renewcommand{\arraystretch}{1.22}
\begin{tabular}{p{0.28\textwidth}p{0.44\textwidth}p{0.19\textwidth}}
\hline
Uncertainty source & Correction or control & Relative standard uncertainty\\
\hline
Reference-field calibration & Individual \mbox{$^{60}\mathrm{Co}$} calibration & 0.85\%\\
Batch uniformity & Individual calibration within the calibrated linear response range & Negligible\\
TLD response repeatability & Evaluated during calibration & 5.0\%\\
TLD response linearity & Measurements within the calibrated linear range & 5.0\%\\
Dose-rate dependence & Experimental dose rate within the applicable range & Negligible\\
Directional dependence & TLD normal aligned with the beam axis & Negligible\\
Photon-energy dependence & Energy-response correction applied according to calibration procedure & Negligible after correction\\
Temperature dependence & Controlled calibration and irradiation conditions & Negligible\\
Humidity dependence & Controlled calibration and irradiation conditions & Negligible\\
Irradiation-to-readout interval & Same readout interval maintained & Negligible\\
Packaging attenuation & \mbox{$0.28\,\mathrm{mm}$} aluminum wrapping & Negligible\\
RGD-3A reader stability & Standard-light check and sensitivity adjustment & 1.0\%\\
\hline
Combined standard uncertainty \mbox{$u_c$} & Uncorrelated components combined & 7.2\%\\
Expanded uncertainty \mbox{$U$} & Coverage factor \mbox{$k=2$} & 14.4\%\\
\hline
\end{tabular}
\end{table}

\subsection{Radiation-induced acoustic signal}

\begin{figure}[!t]
\centering
\includegraphics[width=\textwidth]{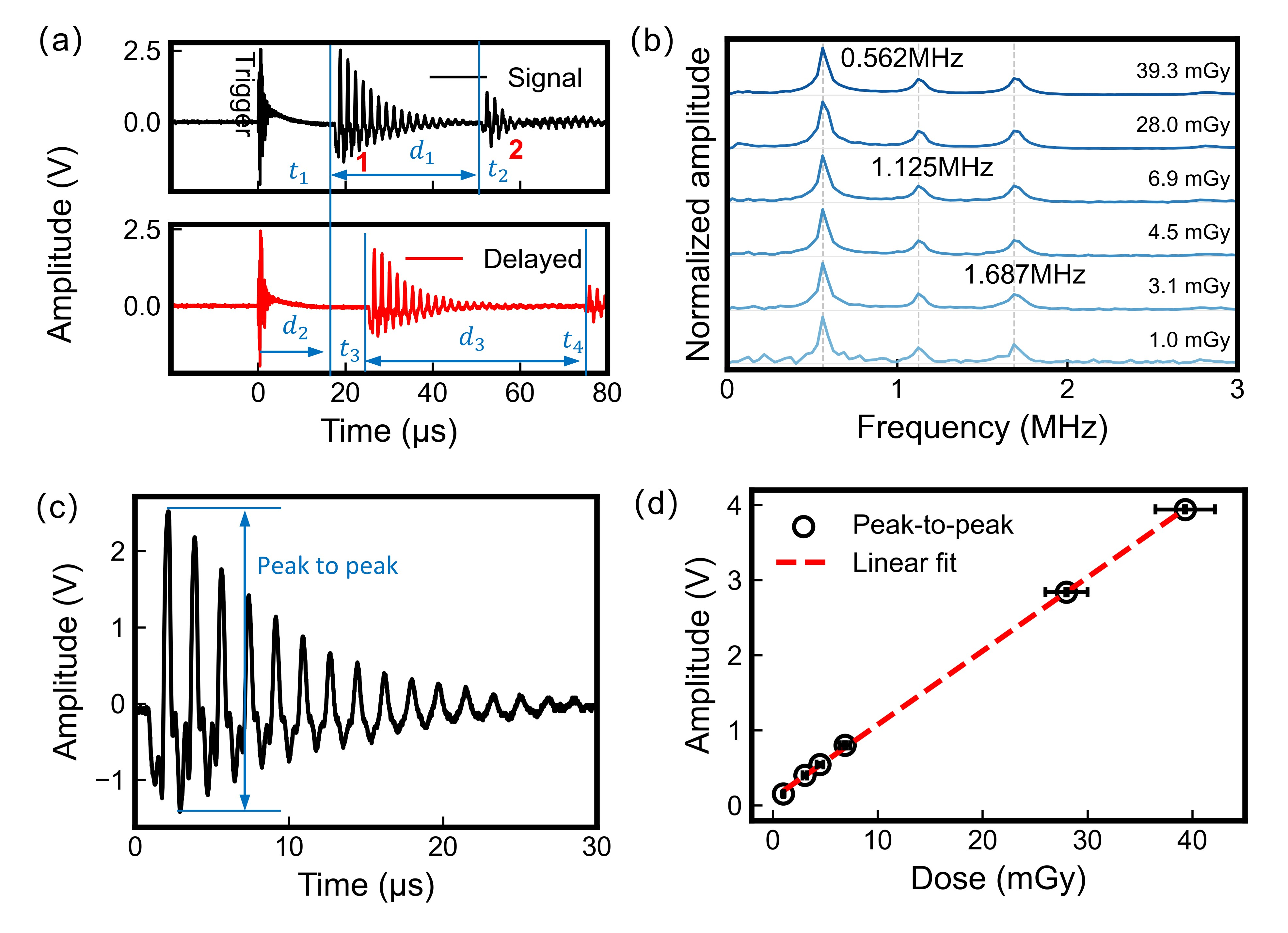}
\caption{(a) Experimentally measured radiation-induced acoustic signal and its delayed signal obtained by increasing the distance between the transducer and the lead foil. (b) Normalized single-sided magnitude spectra of the single-pulse radiation-induced acoustic signals at the six dose levels. Each spectrum is normalized to its own maximum and vertically offset for clarity. The dashed vertical lines mark the approximate positions of the first three principal resonance peaks. (c) Amplified transducer-output voltage corresponding to the measured radiation-induced acoustic signal (component (1), direct pressure wave), with the peak-to-peak definition indicated. (d) Linear regression of the peak-to-peak amplified transducer-output voltage against the TLD-measured single-pulse reference dose \mbox{$D_{\mathrm{ref}}$} [\mbox{$\mathrm{mGy(Si)}$}]. The vertical error bars arise from the baseline noise of the measured acoustic signal, while the horizontal error bars represent the relative combined standard uncertainty of the TLD reference dose (\mbox{$\pm7.2\%$}, \mbox{$k=1$}). Neither set of error bars represents shot-to-shot variability or overall measurement repeatability. Some error bars are smaller than the symbol size and can be seen inside the empty symbols.}
\label{fig:3}
\end{figure}

\mbox{\figsubref{fig:3}{a}} shows a representative time-domain waveform measured in the experiment. The prompt electromagnetic transient near the trigger time (\mbox{$0\mus$}) decays to the background level within approximately \mbox{$10\mus$}. The first acoustic signal, component (1), arrives at \mbox{$t_1=16.7\,\mu\mathrm{s}$} and corresponds to the direct pressure wave generated in the lead foil and transmitted through water to the transducer. Component (2) is an acoustic echo produced after an additional round trip between the transducer and foil surfaces. The temporal separation allows the acoustic signal to be extracted after the electromagnetic transient has decayed to the background level.

To verify the acoustic origin of the signals, the ultrasound transducer was moved farther from the lead foil. In the original configuration, the interval between the direct signal and its echo was \mbox{$t_2-t_1=33.3\,\mu\mathrm{s}$}, consistent with a round trip across the \mbox{$25\,\mathrm{mm}$} water gap at a sound velocity of \mbox{$v_w=1500\,\mathrm{m/s}$}. After the transducer was moved, the direct signal was delayed by \mbox{$t_3-t_1=8\,\mu\mathrm{s}$}, corresponding to an increase of approximately \mbox{$12\,\mathrm{mm}$} in the propagation distance. The interval between the delayed direct signal and its echo was \mbox{$t_4-t_3=49.3\,\mu\mathrm{s}$}, an increase of \mbox{$16\,\mu\mathrm{s}$}, twice the delay of the direct signal. This interval corresponds to an additional round-trip path of approximately \mbox{$74\,\mathrm{mm}$} for the echo relative to the direct signal. These timing relationships are consistent with the expected propagation paths of the direct acoustic wave and its echo, supporting the acoustic origin of the measured signals.

The direct acoustic signal, component (1), exhibits a damped oscillatory waveform, with alternating positive and negative excursions that gradually decrease in amplitude after arrival. The subsequent echo, component (2), also exhibits an oscillatory waveform with a lower amplitude. Within the initial bipolar pulse of component (1), the negative peak has a smaller absolute amplitude than the positive peak. This amplitude asymmetry may reflect contributions from acoustic propagation in the layered structure and the frequency-dependent response of the receiving system, and is further discussed in Section IV through comparison with the layered-medium model.

\mbox{\figsubref{fig:3}{b}} shows the single-sided magnitude spectra obtained by Fourier transforming the single-pulse voltage waveforms at the six dose levels. Each spectrum was normalized to its maximum within \mbox{$0$}--\mbox{$3\,\mathrm{MHz}$}. The first three major spectral peaks occur at approximately \mbox{$0.562$}, \mbox{$1.125$}, and \mbox{$1.687\,\mathrm{MHz}$}. Using the measured longitudinal sound velocity in lead, \mbox{$v_i=2262\,\mathrm{m/s}$}, and foil thickness, \mbox{$l=2.00\,\mathrm{mm}$}, the thickness-resonance relation \mbox{$f_n=nv_i/(2l)$} gives frequencies of \mbox{$0.566$}, \mbox{$1.131$}, and \mbox{$1.697\,\mathrm{MHz}$} for \mbox{$n=1,2,3$}, respectively. The observed peak positions are therefore consistent with the first three thickness resonances of the lead foil.

\subsection{Dose measurements and calibration}

To assess the dose dependence of the measured acoustic response, we first consider the relationship between absorbed dose and initial pressure. The \mbox{$25\,\mathrm{ns}$} X-ray pulse is much shorter than the acoustic transit time across the foil thickness (approximately \mbox{$0.88\,\mu\mathrm{s}$}), satisfying the stress-confinement condition on this scale\cite{ref33}. Neglecting heat diffusion during the pulse, the local initial pressure is expressed as\cite{ref32,ref33}
\begin{equation}
\tag{1}\label{eq:1}
p_0=\Gamma\rho_iD_{\mathrm{Pb}}.
\end{equation}
Here, \mbox{$D_{\mathrm{Pb}}$} is the local absorbed dose in the lead foil, \mbox{$\rho_i$} is the lead density, and \mbox{$\Gamma=\beta v_i^2/C_p$} is the Gr\"uneisen parameter\cite{ref33}, where \mbox{$\beta$} is the volumetric thermal-expansion coefficient and \mbox{$C_p$} is the specific heat capacity at constant pressure\cite{ref34}. For fixed material properties, the local initial pressure is proportional to the local absorbed dose.

The measured acoustic response was quantified by the peak-to-peak voltage, \mbox{$V_{\mathrm{pp}}$}, of the amplified transducer output for the direct signal, component (1), as illustrated in \mbox{\figsubref{fig:3}{c}}. This quantity is the difference between the positive maximum and negative minimum of the direct-signal waveform and is unaffected by a constant baseline offset. For each pulse, \mbox{$V_{\mathrm{pp}}$} was paired with the independently measured TLD reference dose, \mbox{$D_{\mathrm{ref}}$}, expressed in \mbox{$\mathrm{mGy(Si)}$}, to establish the experimental dose-response relationship.

Measurements were performed at six dose levels, with one pulse recorded at each level. The resulting peak-to-peak voltages were fitted against the corresponding TLD reference doses using ordinary least-squares linear regression with a free intercept, as shown in \mbox{\figsubref{fig:3}{d}}. The horizontal and vertical error bars represent the TLD reference-dose uncertainty and the baseline-noise contribution, respectively.

The regression yielded \mbox{$V_{\mathrm{pp}}=0.09803D_{\mathrm{ref}}+0.0940$}, where \mbox{$V_{\mathrm{pp}}$} is in V and \mbox{$D_{\mathrm{ref}}$} is in \mbox{$\mathrm{mGy(Si)}$}. The coefficient of determination was \mbox{$R^2=0.9998$}, with a residual standard error of \mbox{$0.0257\,\mathrm{V}$}, indicating close agreement between the measured responses and the fitted line over the tested dose range. The slope corresponds to a voltage response sensitivity of \mbox{$0.09803\,\mathrm{V/[mGy(Si)]}$} under the present detector and amplification settings, with a 95\% confidence interval of \mbox{$[0.09602,\,0.10004]\,\mathrm{V/[mGy(Si)]}$}.

These results provide an experimental basis for calibrating the detector response against the single-pulse TLD reference dose under the tested conditions. The reported confidence interval characterizes the uncertainty of the fitted voltage-response sensitivity.

The dose-response relationship was established through offline calibration in this study. For prospective online dose estimation, the direct acoustic component can be selected from each acquired waveform, its peak-to-peak voltage extracted, and a TLD-calibrated single-pulse dose estimate obtained using the pre-established calibration relationship, without requiring a new TLD readout for each pulse. In the present setup, the acoustic signal arrives approximately \mbox{$16.7\,\mu\mathrm{s}$} after the X-ray pulse, and the acquisition window is \mbox{$100\,\mu\mathrm{s}$}. The acoustic delay and acquisition window support the feasibility of online operation, while the end-to-end latency of dose estimation, including data transfer and processing, remains to be established.

\section{Theoretical model and analysis}

To interpret the measured resonance spectrum, we develop a one-dimensional layered-medium model that incorporates the depth dependence of energy deposition in the lead foil. Building on the qualitative interface-based description in Section II, the model provides an analytical frequency-domain response for comparison with the experimental spectrum and for examining how the depth-dose distribution affects the relative amplitudes of the resonance peaks.

\subsection{Model assumptions and analytical solution}

The lead foil is modeled as a laterally uniform layer of thickness \mbox{$l$}, occupying \mbox{$-l/2\leq z\leq l/2$}, with density \mbox{$\rho_i$} and longitudinal sound velocity \mbox{$v_i$}. Acoustic propagation is treated as one-dimensional along the thickness direction. Although the experimental foil is bounded by air on the incident side and water on the receiving side, the analytical model approximates the two exterior half-spaces as identical media with the density \mbox{$\rho_o$} and sound velocity \mbox{$v_o$} of water. This approximation exploits the large acoustic impedance contrast between lead and either surrounding medium. X-ray energy deposition is assumed to occur only within the foil.

Within the foil, the acoustic pressure generated by rapid X-ray energy deposition is governed by the one-dimensional thermoacoustic wave equation\cite{ref33}:
\begin{equation}
\tag{2}\label{eq:2}
\left(\frac{\partial^2}{\partial z^2}-\frac{1}{v_i^2}\frac{\partial^2}{\partial t^2}\right)p(z,t)=-\frac{\beta}{C_p}\frac{\partial H(z,t)}{\partial t}.
\end{equation}
Here, \mbox{$p(z,t)$} is the acoustic pressure and \mbox{$H(z,t)$} is the energy deposited per unit volume per unit time. In the exterior media, where energy deposition is neglected, the pressure satisfies the corresponding homogeneous wave equation with sound velocity \mbox{$v_o$}.

The energy-deposition rate is represented by an exponential depth profile and a normalized temporal profile:
\begin{equation}
\tag{3}\label{eq:3}
H(z,t)=Q_0e^{-\mu(z+l/2)}g(t).
\end{equation}
Here, \mbox{$Q_0=\rho_iD_0$} is the deposited-energy density at \mbox{$z=-l/2$}, and \mbox{$D_0$} is the corresponding absorbed dose. The parameter \mbox{$\mu$}, in \mbox{$\mathrm{m^{-1}}$}, controls the direction and steepness of the depth dependence: positive and negative values describe decreasing and increasing deposition with \mbox{$z$}, respectively. The temporal profile satisfies \mbox{$\int_{-\infty}^{+\infty}g(t)\,dt=1$}. Under the short-pulse approximation discussed in Section III-C, \mbox{$g(t)$} is represented by \mbox{$\delta(t)$}.

Using the Fourier convention \mbox{$S(\omega)=\int_{-\infty}^{+\infty}g(t)e^{i\omega t}\,\mathrm{d}t$}, the impulsive temporal profile gives \mbox{$S(\omega)=1$}. Substituting Eq.~\eqref{eq:3} into Eq.~\eqref{eq:2} and transforming in time yields the frequency-domain equation
\begin{equation}
\tag{4}\label{eq:4}
\left(\frac{\mathrm{d}^2}{\mathrm{d}z^2}+k^2\right)p(z,\omega)=
\begin{cases}
\dfrac{i\omega\beta Q_0}{C_p}e^{-\mu(z+l/2)}, & |z|<l/2,\\[4pt]
0, & |z|>l/2.
\end{cases}
\end{equation}
Here, \mbox{$p(z,\omega)$} is the acoustic pressure spectrum, with \mbox{$k=k_i=\omega/v_i$} inside the foil and \mbox{$k=k_o=\omega/v_o$} in the exterior media.

For the exponential source term in Eq.~\eqref{eq:4}, the pressure spectrum inside the foil is the sum of a particular solution associated with energy deposition and two counter-propagating homogeneous waves\cite{ref38}:

\begin{equation}
\tag{5}\label{eq:5}
\begin{aligned}
p_i(z,\omega)&=\frac{i\omega\beta Q_0}{C_p(\mu^2+k_i^2)}e^{-\mu(z+l/2)}\\
&\quad+P_1e^{ik_i z}+P_2e^{-ik_i z}.
\end{aligned}
\end{equation}

The pressure spectra of the waves radiated into the left and right exterior media are, respectively,

\begin{equation}
\tag{6}\label{eq:6}
\begin{aligned}
p_L(z,\omega)&=P_Le^{-ik_o(z+l/2)},\\
p_R(z,\omega)&=P_Re^{ik_o(z-l/2)}.
\end{aligned}
\end{equation}

The coefficients \mbox{$P_1$}, \mbox{$P_2$}, \mbox{$P_L$}, and \mbox{$P_R$} are complex pressure-spectrum amplitudes and are determined by the interface boundary conditions.

Continuity of acoustic pressure and normal particle velocity at the two interfaces, \mbox{$z_L=-l/2$} and \mbox{$z_R=l/2$}, gives\cite{ref38}
\begin{equation}
\tag{7}\label{eq:7}
\begin{aligned}
p_i(z_j,\omega)&=p_j(z_j,\omega),\\
\left.\frac{1}{\rho_i}\frac{\partial p_i}{\partial z}\right|_{z=z_j}
&=\left.\frac{1}{\rho_o}\frac{\partial p_j}{\partial z}\right|_{z=z_j},\quad j=L,R.
\end{aligned}
\end{equation}

Applying the interface conditions to the solutions in Eqs.~\eqref{eq:5} and \eqref{eq:6} gives the pressure spectrum on the receiving side. Let \mbox{$L_w=z-l/2$} denote the propagation distance from the right foil interface, and let \mbox{$\rho=\rho_i/\rho_o$} and \mbox{$c=v_i/v_o$} denote the density and sound-speed ratios, respectively. The resulting spectrum is

\begin{equation}
\tag{8}\label{eq:8}
\begin{aligned}
p_R(z,\omega)&=\frac{i\omega\beta Q_0 A}{C_p[\mu^2+(\omega/v_i)^2]}\\
&\quad\times\Bigl[2\rho c\cos(\omega l/v_i)\\
&\qquad-i[1+(\rho c)^2]\sin(\omega l/v_i)\Bigr]^{-1}\\
&\quad\times e^{i\omega L_w/v_o}.
\end{aligned}
\end{equation}

where the source-coupling factor \mbox{$A$} is given by

\begin{equation}
\tag{9}\label{eq:9}
\begin{aligned}
A
&=
-\frac{i\mu v_{i}}{\omega}
-\rho c
\\
&\quad
+e^{-\mu l}\rho c\cos\left(\frac{\omega l}{v_{i}}\right)
\\
&\quad
+e^{-\mu l}\rho c\frac{\mu v_{i}}{\omega}\sin\left(\frac{\omega l}{v_{i}}\right)
\\
&\quad
-ie^{-\mu l}\sin\left(\frac{\omega l}{v_{i}}\right)
\\
&\quad
+ie^{-\mu l}\frac{\mu v_{i}}{\omega}\cos\left(\frac{\omega l}{v_{i}}\right)
\end{aligned}
\end{equation}

The denominator in Eq.~\eqref{eq:8} describes the resonant response of the layer, while the depth-dependent source distribution influences the excitation of the resonance peaks through the source-dependent terms. The factor \mbox{$e^{i\omega L_w/v_o}$} accounts for propagation through water to the receiving position.

The resonance condition follows from the interface-dependent denominator in Eq.~\eqref{eq:8}, whose magnitude is minimized when
\begin{equation}
\tag{10}\label{eq:10}
\sin\left(\frac{\omega l}{v_i}\right)=0.
\end{equation}
The corresponding resonance frequencies are
\begin{equation}
\tag{11}\label{eq:11}
f_n=\frac{nv_i}{2l},\qquad n=1,2,3,\ldots.
\end{equation}
This is the half-wavelength standing-wave condition for the layer\cite{ref39}. The sound velocity and foil thickness determine these resonance frequencies, while the depth-dose distribution influences the relative excitation of the modes.

For comparison with the measured time-domain waveform, the acoustic pressure at the receiving position is reconstructed from the complex pressure spectrum by inverse Fourier transformation:

\begin{equation}
\tag{12}\label{eq:12}
P_R(z,t)=\frac{1}{2\pi}\int_{-\infty}^{+\infty}p_R(z,\omega)e^{-i\omega t}\,d\omega.
\end{equation}

The interface conditions account for acoustic radiation from the lead foil into the exterior half-spaces. Material attenuation, scattering, nonideal coupling losses, and the frequency-dependent transfer functions of the transducer and amplifier are not included.

\subsection{Model--experiment comparison}

The depth-dose distribution in the lead foil was calculated using TOPAS Monte Carlo simulations\cite{ref36}. The incident bremsstrahlung spectrum was modeled using the prescription of Koch\cite{ref37}, with a mean photon energy of \mbox{$200\,\mathrm{keV}$} and an endpoint energy of \mbox{$1\,\mathrm{MeV}$}. The simulated depth-dose distribution was represented by an exponential profile to determine the parameter \mbox{$\mu$}, which was then used in Eq.~\eqref{eq:8} to calculate the acoustic pressure spectrum. The parameters used for the model--experiment comparison are summarized in Table~\mbox{\ref{tab:1}}.

\begin{table}[!t]
\caption{Parameters used for the comparisons in Fig. 4(a)--(c).}
\label{tab:1}
\centering
\begin{tabular}{p{0.38\linewidth}p{0.52\linewidth}}
\hline
Symbol / Value & Description \\
\hline
$\mu=-764~\mathrm{m}^{-1}$ & Signed depth-dose gradient \\
$\rho_i=11340~\mathrm{kg/m^3}$ & Density of lead \\
$\rho_o=1000~\mathrm{kg/m^3}$ & Density of water \\
\mbox{$v_i=2262~\mathrm{m/s}$} & Pb longitudinal sound velocity measured by pulse echo \\
\mbox{$v_o=1500~\mathrm{m/s}$} & Sound velocity in water \\
\mbox{$l=2.00~\mathrm{mm}$} & Measured lead foil thickness \\
\mbox{$L_w=25~\mathrm{mm}$} & Water propagation distance \\
$\beta=8.34\times10^{-5}~\mathrm{K}^{-1}$ & Volumetric thermal-expansion coefficient of lead \\
$C_p=123~\mathrm{J/(kg\cdot K)}$ & Specific heat capacity of lead at constant pressure \\
\hline
\end{tabular}
\end{table}

\begin{figure}[!t]
\centering
\includegraphics[width=\textwidth]{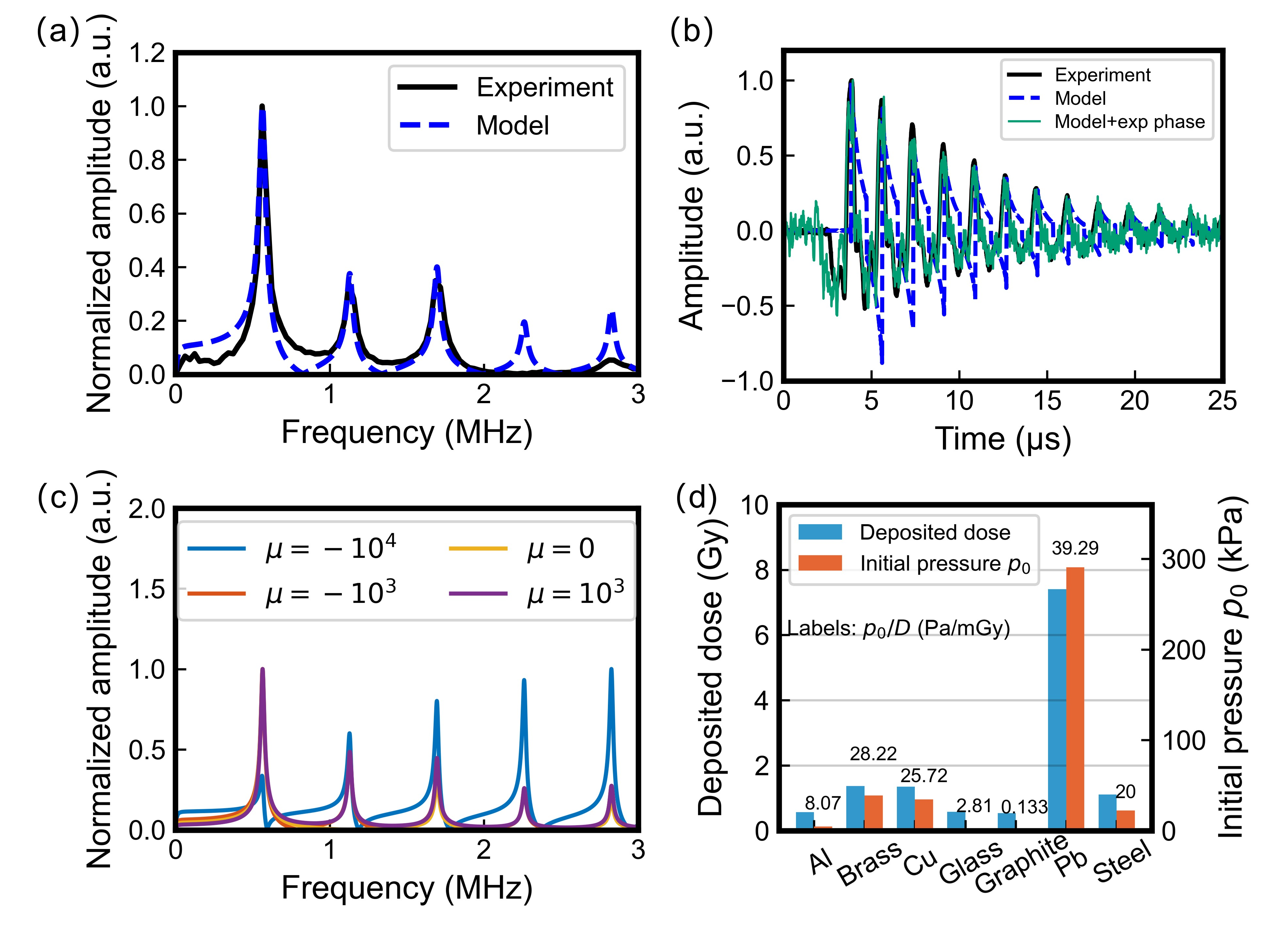}
\caption{(a) Independently normalized experimental (black) and model (blue) magnitude spectra over \mbox{$0$}--\mbox{$3\,\mathrm{MHz}$}. (b) Experimental waveform (black), model waveform (blue dashed), and model magnitude with experimental phase (green). The experimental signal was cropped to \mbox{$15$}--\mbox{$47\,\mu\mathrm{s}$} and demeaned; time is relative to the start of this window. Model and experimental main peaks are aligned before phase replacement. Each waveform is independently max-absolute normalized over \mbox{$0$}--\mbox{$25\,\mu\mathrm{s}$}. The green curve is a same-record phase diagnostic. (c) Frequency spectra for a range of signed depth-dose gradients $\mu$ [$\mathrm{m^{-1}}$]. (d) Deposited dose (blue column) and estimated initial pressure (orange column) for different materials, with the initial-pressure-to-dose coefficients indicated by numerical labels.}
\label{fig:4}
\end{figure}

\mbox{\figsubref{fig:4}{a}} compares the magnitude spectra of the measured voltage and modeled acoustic pressure, each normalized to its maximum within \mbox{$0$}--\mbox{$3\,\mathrm{MHz}$}. Both spectra exhibit a dominant first resonance peak and two weaker peaks below \mbox{$2\,\mathrm{MHz}$}. The experimental peaks occur at approximately \mbox{$0.562$}, \mbox{$1.125$}, and \mbox{$1.687\,\mathrm{MHz}$}, compared with \mbox{$0.564$}, \mbox{$1.130$}, and \mbox{$1.696\,\mathrm{MHz}$} in the model. The model captures the positions and relative amplitudes of these principal peaks, although it predicts stronger resonance peaks above \mbox{$2\,\mathrm{MHz}$} than those observed experimentally. The root-mean-square difference between the normalized spectra over \mbox{$0$}--\mbox{$3\,\mathrm{MHz}$} is \mbox{$0.0605$}. These results support the model's description of the principal resonance features in the normalized spectrum.

The corresponding time-domain pressure waveform was obtained by inverse Fourier transformation of the complex model spectrum using Eq.~\eqref{eq:12}. \mbox{\figsubref{fig:4}{b}} compares this waveform (blue dashed line) with the measured voltage waveform (black line), with each waveform independently normalized to its maximum absolute amplitude. Both exhibit damped oscillations with comparable principal periods and decay trends. However, the model predicts relatively stronger negative peaks, and differences remain in the positive-to-negative peak balance and local waveform shape.

To examine the role of spectral phase in these waveform differences, we retained the model magnitude spectrum and replaced its phase with that of the same experimental record:
\begin{equation}
\tag{13}\label{eq:13}
P_{\mathrm{swap}}(\omega)=|p_R(z,\omega)|e^{i\phi_{\mathrm{exp}}(\omega)},
\end{equation}
where \mbox{$\phi_{\mathrm{exp}}(\omega)$} is the phase of the measured voltage spectrum. Inverse Fourier transformation gives the reconstructed waveform shown by the green line in \mbox{\figsubref{fig:4}{b}}. In the normalized waveform comparison, the reconstruction more closely follows the measured waveform than the direct model prediction, including the relative amplitudes of the positive and negative peaks. The root-mean-square difference between the normalized waveforms over \mbox{$0$}--\mbox{$25\,\mu\mathrm{s}$} decreases from \mbox{$0.3220$} to \mbox{$0.0909$}. Within the \mbox{$2$}--\mbox{$8\,\mu\mathrm{s}$} comparison window, the negative-to-positive peak ratio changes from \mbox{$-0.8809$} to \mbox{$-0.5662$}, approaching the experimental value of \mbox{$-0.5209$}. This reconstruction examines the role of spectral phase using the phase of the same experimental record and therefore does not constitute an independent validation of the model.

Replacing the model phase with the experimental phase improves agreement with the measured waveform, including its positive--negative peak asymmetry. This improvement indicates that spectral phase differences account for part of the discrepancy between the modeled and measured waveforms. The receiving-chain transfer functions and acoustic losses not included in the model may contribute to these waveform discrepancies.

\subsection{Effects of depth-dose distribution and material choice}

The model enables a systematic investigation of how the depth-dose distribution affects the resonant acoustic response. As shown in \figsubref{fig:4}{c}, varying the depth-profile parameter $\mu$ mainly alters the relative amplitudes of the resonant peaks, while the resonant structure itself is preserved. This indicates that the modal response is governed not only by the cavity properties of the foil, but also by how efficiently the thermoacoustic source couples to different eigenmodes. When $\mu=0$, the deposited energy is spatially uniform across the foil thickness and symmetric about the foil center. In this limit, the source term is even in $z$, and therefore couples efficiently only to the symmetric odd-order modes, whereas the antisymmetric even-order modes are suppressed. When $\mu$ is nonzero, the source is no longer symmetric about the foil center, and the even-order modes become allowed.

The sign of $\mu$ determines which interface the deposited energy is biased toward: $\mu>0$ corresponds to energy deposition concentrated toward the X-ray-incident boundary, whereas $\mu<0$ corresponds to deposition concentrated toward the lead-water interface, i.e., the detection side. The magnitude $|\mu|$ characterizes the steepness of the deposited-energy variation across the foil thickness. As $|\mu|$ increases, the depth-dose profile becomes increasingly asymmetric and more strongly localized near one boundary. As a result, the source distribution overlaps with a broader set of cavity eigenmodes, leading to a redistribution of modal amplitudes and a progressive enhancement of higher-order resonances. In the limit of large $|\mu|$, the excitation approaches a boundary-localized impulsive source, yielding a broader modal response with resonant peaks of more comparable amplitudes.

These model results indicate that the RA spectral response contains quantitative information on the spatial distribution of absorbed energy. In particular, the modal amplitude pattern may help constrain the depth-dose profile within the foil.

We further investigated the acoustic responses of different materials. Absorbed dose distributions were obtained from TOPAS Monte Carlo simulations and, together with the material parameters, were used to estimate the expected initial pressure. The pressure values in Fig. 4(d) are initial-pressure estimates calculated using Eq.~\eqref{eq:1}. The materials evaluated included lead, aluminum, steel, copper, brass, glass, and graphite. As shown in \mbox{\figsubref{fig:4}{d}}, the deposited dose varies among the evaluated materials because of differences in their ability to attenuate X-rays. Lead exhibits the highest dose, owing to its high atomic number and density, which yield a large photoelectric absorption cross-section. Brass, copper, and glass show intermediate doses, whereas aluminum and graphite yield the lowest doses because their low Z composition leads to weak X-ray absorption.

The initial pressure is determined not only by the absorbed dose but also by the thermophysical properties of the material. This effect is reflected in the pressure-to-dose ratio ($\mathrm{Pa/mGy}$) indicated in \figsubref{fig:4}{d}. Among the materials compared, lead has the highest initial-pressure-to-dose coefficient of \mbox{$39.29\,\mathrm{Pa/mGy}$}. In contrast, graphite has an extremely small coefficient ($0.133\,\mathrm{Pa/mGy}$), explaining its negligible pressure. Based on the above analysis, lead can be identified as an effective material for radiation-induced acoustic generation.

\section{Conclusions}

This study demonstrates the proof of principle of a resonant radiation-induced acoustic method for dosimetry of ultra-high-dose-rate pulsed X-rays. The main contributions are summarized as follows.

\begin{enumerate}
\item A dosimetry system based on the resonant acoustic response of a lead foil coupled to a passive ultrasonic transducer is established, enabling the response to individual X-ray pulses to be recorded without waveform averaging.
\item Six single-pulse measurements performed with the ``Chen Guang'' accelerator, one at each dose level, demonstrate a linear relationship between the peak-to-peak amplified transducer-output voltage and the TLD reference dose over approximately 1--40 mGy(Si) per pulse. The instantaneous dose rates reached the order of $10^{6}\,\mathrm{Gy/s}$ at the highest dose settings.
\item An analytical frequency-domain solution for radiation-induced acoustic waves in a one-dimensional layered medium is developed. The model captures the principal resonance peak positions and relative amplitudes in the normalized spectral comparison and clarifies how the depth-dose distribution affects the relative excitation of the resonance modes, providing a basis for interpreting the measured spectral response and guiding detector design.
\end{enumerate}

Unlike TLDs and radiochromic films, which require post-irradiation readout, the present method provides single-pulse acoustic signals that could support online dose estimation using a pre-established calibration. Alongside established online approaches based on ionization chambers, semiconductor detectors, or scintillators, it offers an acoustic readout in which the dose-related signal is temporally separated from prompt electromagnetic interference. This separation facilitates signal extraction in pulsed-radiation environments.

The dose-response relationship established here applies to the irradiation conditions and detector configuration used in this study. With other source parameters held fixed, variations in electron-beam charge are expected to affect the acoustic amplitude primarily through changes in X-ray fluence and deposited dose. Changes in accelerating voltage or electron energy may alter the X-ray spectrum and the depth-dose profile in the foil. The model predicts that such profile changes redistribute the relative amplitudes of the resonance modes, while their frequencies remain primarily governed by foil thickness and sound velocity. Pulse-duration variations may also affect the acoustic response, particularly when the stress-confinement approximation is no longer valid. Temperature variations may influence resonance frequencies and dose sensitivity through changes in the foil's thermophysical properties and acoustic coupling. These dependencies should be considered when assessing the applicability of the calibration under different operating conditions.

Measurement repeatability and the overall uncertainty of the proposed dosimetry method remain to be established. Further validation should address these aspects, together with long-term stability and performance across a broader range of irradiation and temperature conditions. Integration of the foil, acoustic coupling layer, and transducer into a compact detector head, together with dedicated amplification and acquisition electronics, could facilitate practical single-pulse dose monitoring. Future studies could also explore the response of this method to other pulsed radiation sources and at higher instantaneous dose rates.

\section*{Acknowledgments}

This work was supported by the National Key Research and Development Program of China (Grant No. 2024YFF0726304), National Key Laboratory of Intense Pulsed Radiation Simulation and Effect Foundation under Grant NKLIPR2423, and the National Natural Science Foundation of China (Grant Nos. 12205008 and 12505306). W.~Ma acknowledges support from the National Science Fund for Distinguished Young Scholars (Grant No. 12225501). The numerical simulations were carried out on the High-Performance Computing Platform of Peking University.

\end{document}